\documentclass[sigplan,10pt, nonacm]{acmart}
\usepackage{xcolor}
\usepackage{xspace}
\usepackage{subcaption}
\usepackage{pifont}
\usepackage{times}
\usepackage{graphbox}
\usepackage{graphicx}
\usepackage{epstopdf}
\usepackage{dblfloatfix} 
\usepackage[htt]{hyphenat}

\usepackage{fancyhdr}
\usepackage{setspace}
\usepackage{relsize}
\usepackage{textcomp}
\usepackage{circledsteps}
\usepackage{url}
\usepackage[super]{nth}
\usepackage{comment}
\usepackage{transparent}
\usepackage{wrapfig}
\usepackage{rotating}

\usepackage{CJK}
\usepackage[utf8]{inputenc}
\usepackage[T1]{fontenc}
\usepackage{soul}
\usepackage{kotex}
\usepackage{lipsum}
\usepackage[normalem]{ulem}
\makeatletter
\def\SOUL@hlpreamble{%
\setul{\dimexpr\dp\strutbox-2pt}{\dimexpr\ht\strutbox+\dp\strutbox-2pt\relax}
\let\SOUL@stcolor\SOUL@hlcolor
\SOUL@stpreamble
}
\makeatother

\newcommand\khlc[1][yellow]{
  \bgroup
  \markoverwith{\textcolor{#1}{\rule[-.5ex]{1pt}{2.5ex}}}
  \ULon
}

\newcommand{\todo}[1]{\bgroup\color{white}\textbf{\khlc[black]{TODO: [#1]}}\egroup\xspace}
\newcommand{\fixme}[1]{\bgroup\color{red}\textbf{\khlc{FIXME: [#1]}}\egroup\xspace}
\newcommand{\pointer}[1]{\bgroup\color{white}\textbf{\khlc[red]{POINTER: [#1 is working here]}}\egroup\xspace}

\newcommand{\reviewer}[1]{\bgroup\color{blue}#1\egroup\xspace}
\newcommand{\ranswer}[1]{\bgroup\color{red}#1\egroup\xspace}

\usepackage{enumitem}
\usepackage[framemethod=tikz]{mdframed}
\tikzset{
    vertical align/.style={
        baseline=-.5*(height("$+$")-depth("$+$"))
    }
}

\usepackage{calc}
\usepackage{amsmath,algorithm,algpseudocode}
\usepackage{mathtools}
\usepackage{cases}
\algrenewcommand\algorithmicindent{0.5em}
\makeatletter
\algrenewcommand\ALG@beginalgorithmic{\footnotesize}
\makeatother
\makeatletter
\renewcommand{\Function}[2]{%
  \csname ALG@cmd@\ALG@L @Function\endcsname{#1}{#2}%
  \def\jayden@currentfunction{#1}%
}
\newcommand{\funclabel}[1]{%
  \@bsphack
  \protected@write\@auxout{}{%
    \string\newlabel{#1}{{\jayden@currentfunction}{\thepage}}%
  }%
  \@esphack
}
\makeatother

\usepackage{tabularx}
\usepackage{array}
\usepackage{multirow}
\usepackage{booktabs}
\usepackage[para]{threeparttable}
\usepackage{colortbl}
\usepackage{multirow}

\makeatletter
\def\hlinewd#1{%
\noalign{\ifnum0=`}\fi\hrule \@height #1 %
\futurelet\reserved@a\@xhline}
\makeatother
\usepackage{hhline}
\newcolumntype{C}{>{\centering\arraybackslash}X}

\usepackage[framemethod=tikz]{mdframed}
\usepackage{marginnote}
\usepackage{pdfcomment}
\mdfsetup{
  hidealllines=true,
  innerleftmargin=2pt,
  innerrightmargin=2pt,
  innertopmargin=2pt,
  innerbottommargin=2pt,
  leftmargin=-2pt,
  rightmargin=-2pt,
  skipabove=-2pt,
  skipbelow=-2pt,
  backgroundcolor=blue!20}

\newlength{\markerHeight}
\newlength{\markerMargin}
\newlength{\linespace}
\newlength{\linedepth}

\sbox0{yf}
\definecolor{mylime}{RGB}{205, 220, 57}
\definecolor{mygreen}{RGB}{60, 200, 0}
\definecolor{myblue}{RGB}{0, 51, 204}

\colorlet{soulred}{red!20}
\colorlet{soulgreen}{green!20}
\colorlet{soulblue}{blue!20}

\usepackage{enumitem}
\setlistdepth{5}
\renewlist{itemize}{itemize}{5}
\setlist[itemize,1]{label=$\bullet$}
\setlist[itemize,2]{label=$\circ$}
\setlist[itemize,3]{label=$\ast$}
\setlist[itemize,4]{label=-}
\setlist[itemize,5]{label=$\cdot$}

\makeatletter
\def\SOUL@hlpreamble{%
\setul{\dimexpr\dp\strutbox-2pt}{\dimexpr\ht\strutbox+\dp\strutbox-2pt\relax}
\let\SOUL@stcolor\SOUL@hlcolor
\SOUL@stpreamble
}
\makeatother

\begin{document}

\acmYear{2025}\copyrightyear{2025}
\setcopyright{acmlicensed}
\acmConference[DIMES '25]{Workshop on Disruptive Memory Systems}{October 13--16, 2025}{Seoul, Republic of Korea}
\acmBooktitle{Workshop on Disruptive Memory Systems (DIMES '25), October 13--16, 2025, Seoul, Republic of Korea}
\acmDOI{10.1145/3764862.3768173}
\acmISBN{979-8-4007-2226-4/25/10}

\title[ScalePool]{ScalePool: Hybrid XLink-CXL Fabric for Composable Resource Disaggregation in Unified Scale-up Domains}

\author{\textls[-8]{Hyein Woo, Miryeong Kwon, Jiseon Kim, Eunjee Na, Hanjin Choi, Seonghyeon Jang, Myoungsoo Jung}}
\affiliation{%
  \institution{Panmnesia, Inc.\\
  \href{https://panmnesia.com/}{https://panmnesia.com}}
  \city{}
  \country{}
}
\email{}

\renewcommand{\shortauthors}{Woo et al.}

\begin{abstract}
 This paper proposes ScalePool, a novel cluster architecture designed to interconnect numerous accelerators using unified hardware interconnects rather than traditional long-distance networking. ScalePool integrates Accelerator-Centric Links (XLink) and Compute Express Link (CXL) into a unified XLink-CXL hybrid fabric. Specifically, ScalePool employs XLink for intra-cluster, low-latency accelerator communication, while using hierarchical CXL-based switching fabrics for scalable and coherent inter-cluster memory sharing. By abstracting interfaces through CXL, ScalePool structurally resolves interoperability constraints, enabling  heterogeneous cluster operation and composable resource disaggregation.

In addition, ScalePool introduces explicit memory tiering: the latency-critical tier-1 combines accelerator-local memory with coherence-centric CXL and XLink, whereas the high-capacity tier-2 employs dedicated memory nodes interconnected by a CXL-based fabric, achieving scalable and efficient memory pooling. Evaluation results show that ScalePool accelerates LLM training by 1.22$\times$ on average and up to 1.84$\times$ compared to conventional RDMA-based environments. Furthermore, the proposed tier-2 memory disaggregation strategy reduces latency by up to 4.5$\times$ for memory-intensive workloads.
\end{abstract}



\maketitle

\section{Introduction}
With the rapid growth of AI workloads, the volume of data that must be processed and preserved has increased exponentially. In parallel, modern large-scale AI workloads, such as large language models (LLMs) \cite{vaswani2017attention,brown2020language,rae2021scaling,grattafiori2024llama,chowdhery2023palm,shoeybi2019megatron} and recommendation systems \cite{naumov2019deep,ardestani2022supporting,mudigere2022software}, demand a diverse range of performance metrics beyond mere computational throughput, often requiring compute capacities and memory footprints that significantly exceed the capabilities of individual accelerators. As a result, it has become challenging for a single hardware type to efficiently support various AI applications. To meet these expanding requirements, data centers comprising thousands to tens of thousands of homogeneous or heterogeneous accelerators have been constructed, enabling the handling of diverse application services. These data centers typically follow a hierarchical organization, starting from nodes, which is the smallest computing units, aggregated into racks, with multiple racks forming rows.

Hyperscalers integrate these nodes, racks, and rows into modular data center units at the building scale and manage multiple such sites globally. To interconnect accelerators and modularize data centers, two primary scaling methods have been employed: \textit{scale-up} and \textit{scale-out}. Scale-up architectures rely on high-speed hardware interconnects, enabling connected devices to operate as a unified system but supporting only a limited number of accelerators. In contrast, scale-out architectures mitigate this limitation by leveraging long-distance network technologies, such as InfiniBand and Ethernet, allowing a larger number of accelerators to be interconnected for distributed and parallel processing.

Although inter-device connectivity in both scale-up and scale-out configurations has improved in terms of bandwidth and latency, the boundary between these two approaches remains unchanged. In practice, scale-up configurations are restricted to connecting tens of accelerators, resulting in most data communication, such as tensor exchanges or gradient synchronizations, occurring within scale-out domains in current AI infrastructures employing thousands to tens of thousands of accelerators. Despite the high-bandwidth capabilities in scale-out networks, their performance falls short of that of scale-up architectures. During the data transfers, in particular, software interventions are inevitable. Even performance-optimized frameworks such as RDMA cannot completely eliminate performance degradation due to unnecessary data copying across different computing domains, serialization/deserialization, and computational overhead.

In this paper, we propose ScalePool, a large-scale cluster architecture designed to interconnect numerous accelerators through hardware-based interconnects rather than relying on long-distance network technologies. The primary goal of ScalePool is to mitigate the performance degradation inherent in scale-out configurations by integrating diverse accelerator and memory interconnect technologies, thereby enabling scalable yet composable resource disaggregation within a large-scale, unified scale-up architecture. Specifically, ScalePool employs a unified XLink-CXL interconnect, referred to as a \textit{hybrid fabric}, which combines the strengths of \emph{Compute Express Link} \cite{cxl} and \emph{Accelerator-Centric Links} (XLink), including Ultra Accelerator Link (UALink) \cite{ual} and NVIDIA’s NVLink; CXL supports coherent memory pooling and resource disaggregation across clusters, but it can be suboptimal for latency-sensitive accelerator communication at large-scale. In contrast, XLink excels in accelerator-to-accelerator communication with low overhead and high bandwidth but lacks scalability and coherent memory support.

\begin{table}[t]
    \centering
    \renewcommand{\arraystretch}{1.05}
    \resizebox{1.03\columnwidth}{!}{
    \begin{tabular}{|l|l|l|l|}
    \hline
    \textbf{Feature} & \textbf{CXL} & \textbf{UALink} & \textbf{NVLink} \\
    \hline
    Main purpose & Memory sharing & Accelerator comm. & Accelerator comm. \\
    Latency & Medium (ns) & Low (sub-µs) & Very low (ns) \\
    Coherence & Cache-coherent & Non-coherent & Limited coherence \\
    Topology & Flexible fabric & Single-hop & Single-hop \\
    Compatibility & Open standard & Vendor-neutral & NVIDIA-centric \\
    PHY & PCIe-based & Ethernet-based & Proprietary \\
    \hline
    \end{tabular}}
    \vspace{8pt}
    \caption{Key Differences among CXL, UALink, and NVLink.}
    \label{tab:cxl_ualink_nvlink_comparison}
    \vspace{-12pt}
\end{table}

Thus, the proposed ScalePool employs XLink for high-bandwidth intra-cluster accelerator interactions while leveraging hierarchical CXL-based switching fabrics for scalable and coherent inter-cluster memory sharing. In addition, ScalePool structurally overcomes interoperability constraints between NVLink and UALink clusters by abstracting interfaces through CXL, enabling independent and heterogeneous cluster operation alongside efficient inter-cluster memory sharing and data movement. ScalePool is further optimized to address diverse memory performance and capacity requirements through explicit memory tiering. The latency-critical tier-1 memory layer combines accelerator-local high-speed memory with lightweight coherence-centric CXL and XLink to ensure superior performance. The high-capacity tier-2 memory layer employs dedicated memory nodes connected via a CXL-based fabric, excluding accelerators and CPUs, thereby providing an efficiently-scalable memory pool.

Our evaluation results show that the proposed ScalePool architecture and hierarchical memory approach achieve an average speedup of 1.22$\times$ and a maximum of 1.84$\times$ for large-scale language model training compared to conventional RDMA-based environments. Furthermore, the resource disaggregation strategy using the tier-2 memory pool significantly reduces latency by up to 4.5$\times$ for memory-intensive workloads.

\section{Background and Motivation}
Contemporary AI workloads exhibit distinct computational characteristics, which frequently require tensor-level communication \cite{jia2019beyond,narayanan2021efficient,fei2021efficient}, accelerator synchronization \cite{zhang2017poseidon,huang2019gpipe}, and memory-intensive operations, including embedding lookups \cite{zha2022dreamshard,cheng2016wide}, \textit{key-value} (KV) caching \cite{liu2024cachegen,lee2024infinigen,liu2023scissorhands}, and \textit{retrieval-augm\-ented generation} (RAG) \cite{lewis2020retrieval,guu2020retrieval,borgeaud2022improving}. Addressing these diverse requirements with a single hardware design remains challenging. For example, training LLMs demands scalable model management and substantial network bandwidth, whereas inference tasks emphasize computational intensity. Note that KV caching and RAG require extensive memory capacities combined with high I/O bandwidth \cite{kwon2023efficient,adnan2024keyformer}. To meet these simultaneous demands, various interconnect technologies have been proposed to integrate heterogeneous accelerators and different types of memory resources.

Among those interconnect technologies, CXL and XLink have emerged, each exhibiting distinct architectural characteristics. Table~\ref{tab:cxl_ualink_nvlink_comparison} summarizes the differences across these technologies, and their details are explained below.

\begin{figure}
    \centering
    \includegraphics[width=.99\linewidth]{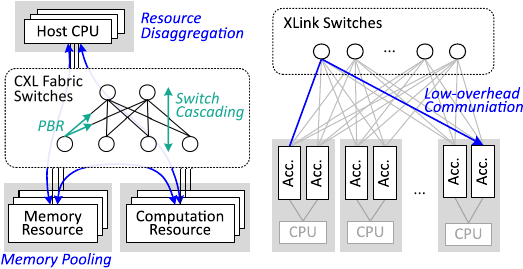}
    \begin{subfigure}{\linewidth}
        \vspace{5pt}
        \begin{tabularx}{\textwidth}{
            p{\dimexpr.47\linewidth-2\tabcolsep-1.3333\arrayrulewidth}
            p{\dimexpr.47\linewidth-2\tabcolsep-1.3333\arrayrulewidth}
            }
            \caption{CXL.} \label{fig:cxl} &
            \caption{XLink.} \label{fig:xlink}
        \end{tabularx}
    \end{subfigure}
    \vspace{-18pt}
    \caption{Architectural differences.} \label{fig:background}
\end{figure}

\noindent \textbf{Compute express link (CXL).} As shown in Figure~\ref{fig:cxl}, CXL particularly targets coherent memory pooling and resource disaggregation within data centers with three sub-protocols: \emph{CXL.mem}, \emph{CXL.cache}, and \emph{CXL.io}. The CXL.mem protocol enables asynchronous memory access by permitting the memory controller, typically embedded within the host CPU, to be located anywhere across the interconnect network \cite{das2024introduction,sun2023demystifying,ahn2022enabling}. CXL.cache manages coherence among multiple computing resources. CXL.io focuses on bulk I/O operations unrelated to memory or cache coherence and resembles PCIe.

Recent versions of the CXL specification introduced advanced features \cite{cxl}, such as \textit{switch cascading} and \textit{port-based routing} (PBR). Specifically, cascading enables multiple switches to interconnect hierarchically, facilitating various network topologies. PBR allows traffic routing decisions to be determined at each switch port, supporting efficient operation of these diverse topologies. Together, these advanced features enable the underlying interconnect to form fabric structures composed of multi-level switches. While this capability distinguishes CXL from other XLink technologies, the cache coherence mechanisms and hierarchical switching structures in CXL may introduce additional latency, in the context of frequent tensor exchanges among accelerators.

\noindent \textbf{Accelerator-centric link (XLink).}
As shown in Figure~\ref{fig:xlink}, XLink technologies, including UALink and NVLink, are designed for low-overhead intra-cluster communication, primarily focusing on accelerator interactions. UALink adopts Ethernet-based PHY interfaces with point-to-point connectivity, providing up to 100 GB/s bandwidth per port at sub-$\mu$s latency. NVLink is specialized for GPU-to-GPU communication, delivering latency below 500 ns and efficient tensor-level data transmission through proprietary signaling schemes. Both links employ a common strategy that prioritizes direct inter-accelerator data transfers; thus, in this paper, we collectively refer to these technologies as XLink. Specifically, XLink employs a single-hop switched topology (one-stage Clos or mesh), facilitating efficient accelerator interconnections through large data flit sizes (640B for UALink and 48B--272B for NVLink). Despite optimization for accelerator-level connections, XLink exhibits two primary limitations. First, its point-to-point, single-hop topology constrains scalability, preventing fabric-level interconnection across numerous devices. Second, XLink does not support cache coherence and memory sharing beyond closely coupled accelerator clusters.

\noindent \textbf{Interoperability limitation.}
UALink emphasizes openness and interoperability, supporting integration of diverse accelerators from multiple vendors. In contrast, NVLink has targeted NVIDIA-centric deployments. The recent \textit{NVLink Fusion} extension addresses this limitation in part by providing coherent external interfaces \cite{NVLinkfusion}. Specifically, NVLink Fusion offers two distinct interfaces: a chip-to-chip (C2C) and a GPU-to-GPU. The C2C interface is open and accessible for third-party usage, whereas the GPU-to-GPU interface remains proprietary but allows third-party integration in a chiplet-based format. Although NVLink Fusion introduces new opportunities to interconnect heterogeneous accelerator architectures, NVIDIA's strategic policy still mandates inclusion of at least one NVIDIA component within NVLink-connected system.

\begin{figure}
    \centering
    \includegraphics{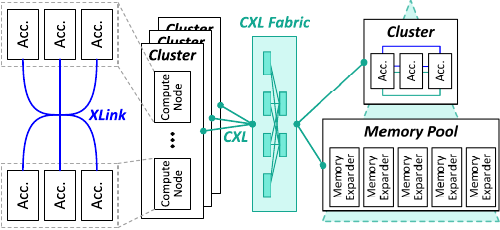}
    \begin{subfigure}{\linewidth}
        \vspace{5pt}
        \begin{tabularx}{\textwidth}{
            p{\dimexpr.47\linewidth-2\tabcolsep-1.3333\arrayrulewidth}
            p{\dimexpr.57\linewidth-2\tabcolsep-1.3333\arrayrulewidth}
            }
            \caption{Accelerator clusters.} \label{fig:scalepool} &
            \caption{Tiered memory.} \label{fig:tiering}
        \end{tabularx}
    \end{subfigure}
    \vspace{-16pt}
    \caption{XLink-CXL hybrid fabric.} \label{fig:cluster_architecture}
    \vspace{-5pt}
\end{figure}

\section{Overview of Hybrid Fabric-based ScalePool}
Combining CXL with XLink provides architectural benefits. Specifically, such hybrid architectures utilize XLink for efficient, low-overhead intra-cluster communication, while leveraging CXL for scalable coherent memory sharing and resource disaggregation. In addition, CXL resolves interoperability constraints between UALink and NVLink, enabling data centers to seamlessly incorporate heterogeneous accelerator devices within a unified, scale-up interconnect domain.

The proposed architecture encompasses two major design concepts: i) ``accelerator-centric clusters'', specialized for efficient intra-cluster accelerator communication, and ii) ``tiered memory architectures'', employing disaggregated memory pools to handle large-scale data management (cf. Figure~\ref{fig:cluster_architecture}). XLink facilitates low-latency and high-bandwidth intra-node communication using direct, single-hop topologies, which are beneficial for bandwidth-intensive operations, including frequent tensor exchanges and gradient synchronization. However, the inherent constraint of single-hop connectivity limits scalability, restricting the number of accelerators and memory devices that can be interconnected within a cluster. In contrast, CXL provides scalable connectivity across multiple clusters or data centers via multi-level cascaded switches, allowing diverse interconnect topologies and coherent memory pools. CXL's capability for dynamic memory allocation addresses critical memory-intensive scenarios, such as KV caching and RAG. Moreover, CXL facilitates efficient inter-node data sharing through protocol-level cache coherence and instruction-level memory transactions, reducing redundant data transfers and improving overall memory utilization.

In this architecture, composable disaggregation physically separates computing resources from memory pools, supporting independent scalability, simplified maintenance, and flexible hardware upgrades. Compute nodes interconnected through XLink, combined with memory resources managed via CXL, provide operational flexibility to swiftly transition between compute-intensive training and latency-sensitive inference workloads. In addition, memory resources can be structured into hierarchical tiers, enabling optimized allocation aligned with workload-specific performance and capacity demands.

\section{Accelerator-Centric Cluster Architecture}
\label{sec:cluster}
Scaling beyond individual accelerator clusters requires efficient inter-cluster communication to support extensive resource sharing and composability across the broader data center infrastructure. In this context, the term ``cluster'' denotes a rack-scale system comprising multiple accelerator nodes. Unlike intra-cluster interconnects, which commonly adopt direct, point-to-point topologies, inter-cluster networks demand more scalable and flexible communication structures. CXL addresses these requirements by employing hierarchical, multi-level switching topologies, facilitating coherent memory pooling among distributed accelerator clusters.

This subsection introduces the hybrid fabric-based architecture called \textit{ScalePool}, designed as a scalable, hierarchical interconnection optimized for accelerator-intensive tasks. ScalePool integrates multiple accelerator clusters interconnected via CXL fabrics. Within each accelerator cluster, XLink serves as the primary intra-cluster interconnect, enabling direct, high-bandwidth, and low-latency communication among accelerators. Detailed cluster configurations are presented below.

\begin{figure}
    \centering
    \includegraphics[width=.97\linewidth]{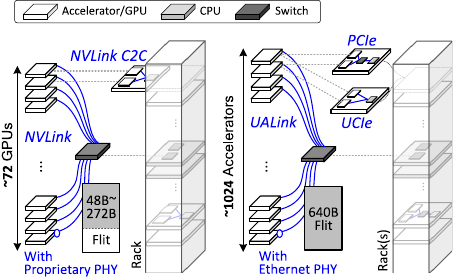}
    \begin{subfigure}{\linewidth}
        \vspace{10pt}
        \begin{tabularx}{\textwidth}{
            p{\dimexpr.45\linewidth-2\tabcolsep-1.3333\arrayrulewidth}
            p{\dimexpr.56\linewidth-2\tabcolsep-1.3333\arrayrulewidth}
            }
            \caption{NVLink cluster.} \label{fig:nvlink} &
            \caption{UALink cluster.} \label{fig:ualink}
        \end{tabularx}
    \end{subfigure}
    \caption{Accelerator clusters.} \label{fig:intracluster}
\end{figure}

\noindent \textbf{Intra-cluster design with XLink.} Within the ScalePool architecture, NVLink and UALink serve as intra-cluster interconnect technologies optimized for accelerator communication. These XLink technologies utilize a point-to-point connection to support intra-cluster accelerator interactions, as illustrated in Figure~\ref{fig:intracluster}. Specifically, NVLink clusters typically consist of GPU-centric nodes interconnected via NVSwitches, accommodating up to 72 GPUs per rack. CPUs within these clusters connect through NVLink C2C interfaces. Similarly, UALink targets accelerator connectivity within clusters or racks, supporting theoretical scales of up to 1,024 accelerators using a single-hop topology. In practical deployments, however, larger accelerators such as GPUs limit UALink cluster scales to configurations similar to NVLink clusters (72 accelerators per rack). CPUs in UALink clusters connect via PCIe switches or short-reach interfaces such as UCIe \cite{ucie2.0}.

Fundamental differences in physical layers (PHY) and data formats restrict combining NVLink and UALink within a single accelerator cluster. As explained earlier, NVLink employs proprietary PHY interfaces with flit sizes ranging from 48B to 272B, whereas UALink adopts Ethernet-based PHY layers with larger, fixed 640B flits. In addition, NVLink's proprietary interface requirement necessitates incorporating NVIDIA components in NVLink-based clusters, thus limiting integration with third-party accelerator configurations.

\begin{figure}
    \centering
    \includegraphics[width=.94\linewidth]{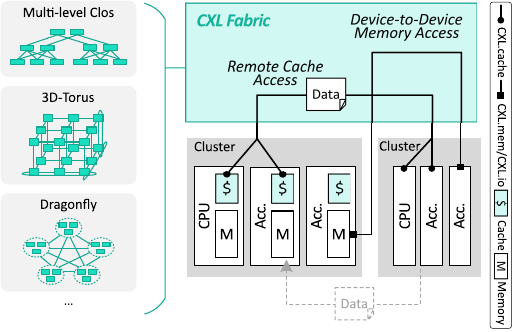}
    \begin{subfigure}{\linewidth}
        \vspace{7pt}
        \begin{tabularx}{\textwidth}{
            p{\dimexpr.37\linewidth-2\tabcolsep-1.3333\arrayrulewidth}
            p{\dimexpr.63\linewidth-2\tabcolsep-1.3333\arrayrulewidth}
            }
            \caption{Fabric topologies.} \label{fig:cxl_topology} &
            \caption{Inter-cluster communication.} \label{fig:cxl_latency}
        \end{tabularx}
    \end{subfigure}
    \caption{ScalePool architecture.} \label{fig:intercluster}
\end{figure}

Therefore, NVLink-based clusters primarily consist of NVIDIA GPUs, complemented by specialized accelerators optimized for tasks poorly suited to GPU architectures. Example workloads include branch-intensive computations, irregular memory access patterns (e.g., graph processing, sparse matrix operations), or latency-sensitive real-time operations. In contrast, UALink-based clusters integrate non-NVIDIA accelerators, such as AMD GPUs, Meta’s MTIA~\cite{coburn2025meta}, Amazon’s Trainium~\cite{trainium}, Inferentia~\cite{inferentia}, Microsoft’s Maia~\cite{maia}, and Intel’s Gaudi~\cite{gaudi}. UALink’s vendor-neutral design facilitates diverse accelerator deployments and supports intra-cluster communication independent of proprietary technologies.

Aligning interconnect technologies and accelerator choices enables the ScalePool architecture to optimize intra-cluster communication, computational throughput, and resource efficiency in heterogeneous accelerator environments.

\begin{figure*}[t]
    \centering
    \includegraphics[width=\textwidth]{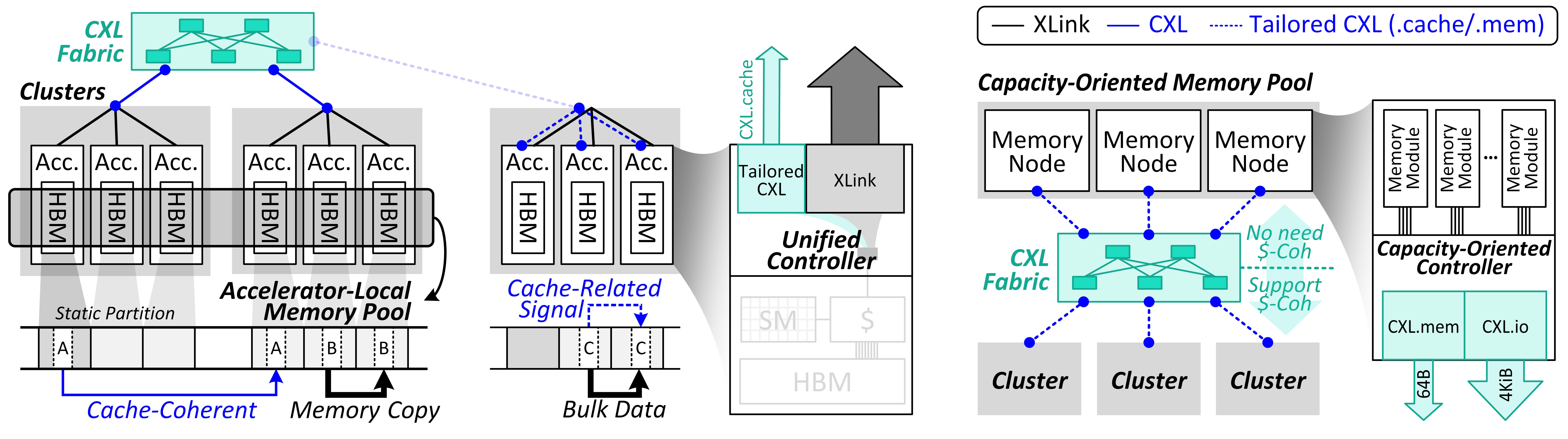}
    \begin{subfigure}{\textwidth}
        \begin{tabularx}{\textwidth}{
            p{\dimexpr.32\linewidth-2\tabcolsep-1.3333\arrayrulewidth}
            p{\dimexpr.28\linewidth-2\tabcolsep-1.3333\arrayrulewidth}
            p{\dimexpr.40\linewidth-2\tabcolsep-1.3333\arrayrulewidth}
            }
            \caption{Accelerator-local memory pool.} \label{fig:tier-1} &
            \caption{Unified controller.} \label{fig:unified_controller} &
            \caption{Capacity-oriented memory pool and controller.} \label{fig:tier-2}
        \end{tabularx}
    \end{subfigure}
    \caption[width=\textwidth]{Tiered memory architectures.} \label{fig:tiered_memory_pool}
\end{figure*}

\noindent \textbf{Inter-cluster design with CXL.} In ScalePool, XLink clusters form a unified hierarchical system via CXL fabrics, creating a large-scale multi-accelerator environment. The hybrid fabric architecture reduces latency and data-transfer overhead by leveraging XLink for intra-cluster accelerator communication and CXL for scalable inter-cluster memory sharing. Unlike XLink, CXL adopts multi-level switch-based architectures, enabling dynamic aggregation of distributed memory resources into composable memory pools. These coherent pools can reduce reliance on off-chip memory or storage, improving resource utilization and flexibility. Furthermore, CXL resolves interoperability constraints between XLink clusters by abstracting each cluster into independent entities and mediating their interactions within a unified architecture.

Figure~\ref{fig:cxl_topology} shows hierarchical CXL-based fabric topologies interconnecting multiple XLink-based accelerator-centric clusters within a unified ScalePool. Leveraging PBR routing and multi-level switch cascading, CXL supports various fabric structures, including multi-level Clos, 3D-Torus, and DragonFly, accommodating different data center requirements. This flexibility enables ScalePool to efficiently adapt to workload-specific needs, integrating diverse accelerators, memories, and computing resources into unified scale-up domains \cite{lerner2024cxl,ahn2024examination,sharma2022compute,chen2024next}.

Another advantage of this hybrid fabric approach is inter-cluster cache coherence. Utilizing CXL.cache, accelerators can directly access remote memory at instruction-level granularity without software involvement, aggregating distributed accelerator-local memories into unified memory spaces. Therefore, accelerators fetch data immediately from remote caches, minimizing conventional inter-node access overhead. In addition, CXL provides separate sub-protocol interfaces for memory access (CXL.mem) and bulk data transfer (CXL.io), supporting various data transfer granularities and enabling direct device-to-device communication without CPU intervention. As shown in Figure~\ref{fig:cxl_latency}, this design reduces redundant data movements among accelerators and enhances computational throughput, especially for memory-intensive workloads.

Finally, protocol-level coherence supported by CXL enables efficient collective communication (e.g., broadcast, scatter/gather, all-reduce) by eliminating explicit synchronization and redundant data copying overhead. Since distributed accelerators share coherent memory pools, hardware implicitly manages data movements, simplifying accelerator kernel development. Software developers thus can focus on computational logic without handling explicit synchronization or data movement instructions. Accelerator-internal caches further optimize data access locality, maximizing computational performance and efficiency across diverse workloads.

\section{Memory Tiers with CXL Optimizations}
\label{sec:tier}
Building upon the hybrid fabric, we propose ScalePool, a scalable architecture that incorporates a tiered memory hierarchy to address the diverse memory-performance demands of contemporary AI workloads. This structure comprises two distinct memory tiers: i) high-performance local memory managed via XLink and \textit{coherence-centric CXL}, and ii) scalable, composable memory pools enabled through \textit{capacity-oriented CXL}. To effectively establish these two memory tiers, we recommend lightweight implementations of CXL specifically tailored for each of these tiers.

\noindent \textbf{Tier-1: Accelerator-local memory pool.}
Accelerator clusters in ScalePool connect via XLink and utilize high-performance memory technologies integrated within each accelerator, including on-package HBM or specialized DDR modules. As ScalePool expands, memory capacities, types, and usage patterns differ across clusters based on workload characteristics. Since ScalePool includes a CXL fabric for inter-cluster connections, as shown in Figure~\ref{fig:tier-1}, these distributed accelerator-internal memory resources across multiple clusters can form a unified coherent memory as the tier-1 memory pool.

Within each cluster, XLink establishes a unified linear memory address space by statically partitioning accelerator-internal memories. For instance, UALink organizes accelerator memories into a NUMA-like domain, and NVLink employs virtualization techniques for similar unification. However, such unified memory lacks protocol-level coherence. Thus, sharing data beyond static partitions requires explicit software-managed copying. Memory access to non-local regions incurs XLink data transfer latency, especially significant across cluster boundaries, impacting performance. To address these issues, CXL can be employed between clusters. Without modifying existing protocols, clusters can designate specific memory regions within accelerators as cache-coherent and expose them to the inter-cluster CXL fabric. This selective coherence approach enables cache-coherent data sharing for targeted datasets or applications, improving data locality and performance. Frequently accessed data remains within accelerator caches, eliminating unnecessary inter-cluster transfers.

For workloads requiring extensive cache coherence, lightweight coherence-centric CXL integration within clusters is feasible. As shown in Figure~\ref{fig:unified_controller}, dedicated CXL coherence logic can be embedded into accelerators alongside existing XLink controllers, enabling fully unified coherent memory across accelerators within clusters. This integration removes the need for explicit data copying or collective operations, enhancing performance. Although adding dedicated coherence logic introduces design complexity, cost, and potential redundant traffic, these issues can be managed by optimizing the CXL protocol specifically for coherence traffic. In practice, bulk data movements occur via XLink, while optimized implementations of CXL.cache handle only coherence transactions. This combined approach improves performance, simplifies data management, and enhances computational efficiency.

Combining accelerator-internal memory via XLink and coherence-oriented CXL addresses low-latency and coherence demands at the accelerator-node level. However, modern AI workloads often require memory capacity exceeding local rack-level resources. To meet these larger memory requirements, scalable composable memory pools, as described in the following section, complement the proposed approach.

\noindent \textbf{Tier-2: Capacity-oriented memory pool.}
Accelerator-local memory serves frequently accessed and performance-critical data. However, modern AI workloads often demand greater memory capacities despite potential performance trade-offs. Examples include embedding tables, large caches, and external knowledge bases. To address these requirements, we propose a tier-2 composable memory structure integrated into ScalePool, enabling flexible capacity expansion.

As illustrated in Figure~\ref{fig:tier-2}, the proposed two-tier architecture consists of memory nodes physically separated from accelerator clusters and interconnected via a dedicated CXL fabric. Within ScalePool, tier-1 accelerator-local memory provides a unified, coherent memory space managed by coherence-centric CXL and XLink controllers. Therefore, accesses to tier-2's memory nodes occur only when workloads exceed the memory capacities available at the rack-level accelerators. Such scenarios traditionally rely on external storage or distributed file systems with millisecond- to second-level latencies. In contrast, tier-2's composable memory nodes reduce these latencies to tens or hundreds of nanoseconds. These memory nodes integrate memory modules, excluding CPUs or accelerators to maximize density and resource efficiency.

Memory nodes can be positioned flexibly within ScalePool, provided connectivity via the CXL fabric is maintained. Node placement, either physical or virtual, is guided by the spatial management policies and latency considerations of the data center. Locating memory nodes closer to accelerator clusters reduces dependency on slower external storage and scale-out access methods. Thus, these nodes of the tier-2 effectively serve as capacity-oriented memory pools explicitly optimized for memory expansion and scalability.

Similar to coherence-centric CXL designs, capacity-oriented CXL implementations may leverage existing hybrid fabrics or optimize further for large-scale deployments. Since tier-1 accelerator-local memory already manages coherence for latency-sensitive data, tier-2 memory nodes can focus solely on capacity, simplifying controller complexity. For instance, cache coherence across tier-2 nodes is unnecessary, enabling selective deactivation of CXL.cache or CXL.io protocols at switches and endpoints to improve cost efficiency. If tier-1 memory exclusively acts as a cache, the tier-2 pool can omit the CXL.mem protocol entirely, relying solely on CXL.io for bulk data transfers. Regardless of the selected approach, adequate CXL fabric ports are essential to maintain optimal data-transfer performance between memory tiers.

\section{Preliminary Evaluation}

\noindent \textbf{Methodology.} To evaluate the proposed ScalePool architecture, we conducted preliminary assessments that model critical performance factors, including link latency derived from flit sizes, PHY layer characteristics, and packetization and queuing behaviors at both link and transaction layers. Switch latencies were determined using empirical measurements from our silicon prototypes, factoring in the hop counts required for endpoint-to-endpoint communication. These modeled latencies were integrated into a high-level LLM co-design simulation framework \cite{isaev2023calculon}. Our baseline architecture employs XLink for intra-rack communication, whereas inter-rack connections across clusters rely on InfiniBand leveraging RDMA-based data transfers. Specifically, our modeled baseline configuration represents a typical NVLink-based cluster consisting of 36 GB200 modules \cite{gb200}, with 72 GPUs interconnected via NVLink 5.0.

\noindent \textbf{Workloads and configurations.}
The evaluation encompasses five transformer-based contemporary LLMs: GPT-3 \cite{brown2020language}, Gopher \cite{rae2021scaling}, Llama 3 \cite{grattafiori2024llama}, PaLM \cite{chowdhery2023palm}, and Megatron \cite{shoeybi2019megatron}. Simulation parameters, including GPU count, parallelism degree, batch size, and applied optimizations, adhere to the configurations originally presented in each model's initial research. In addition, all evaluated scenarios assume the common training optimization methods of weight offloading and optimizer offloading \cite{ren2021zero}. Offloading refers to storing necessary training data in external memory to free GPU-local memory capacity. The baseline scenario places offloaded data in CPU-attached memory of the GB200 modules, while ScalePool employs a dedicated CXL memory pool for external data storage.

\begin{figure}
    \centering
    \includegraphics[width=\linewidth]{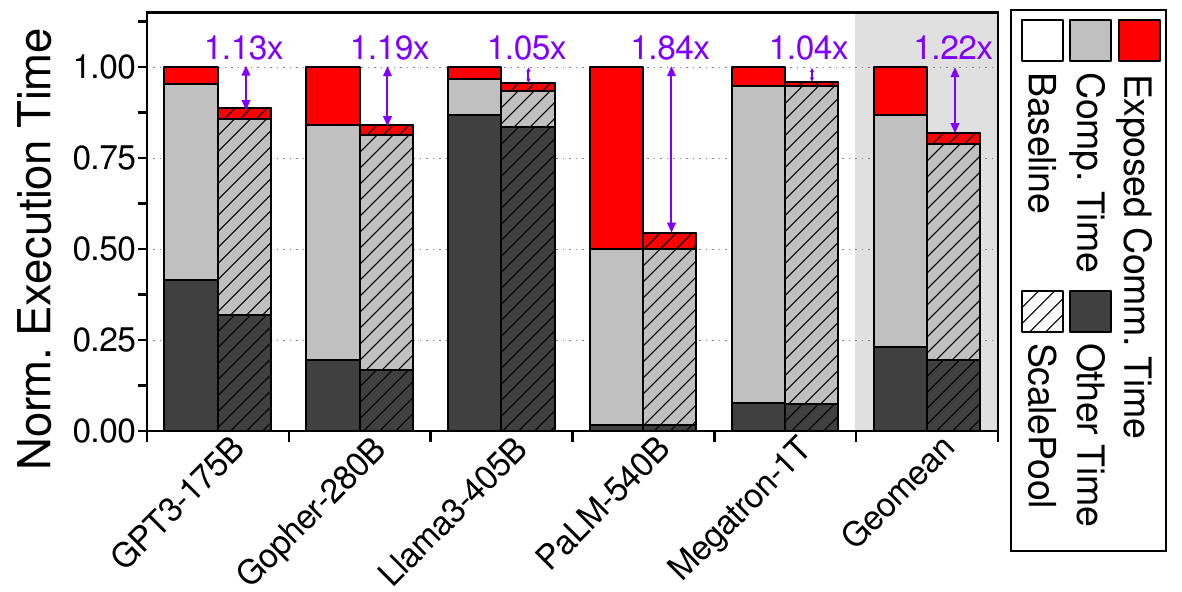}
    \vspace{-10pt}
    \caption{Performance results for ScalePool.} \label{fig:eval_ScalePool}
\end{figure}

Figure~\ref{fig:eval_ScalePool} presents the LLM training execution time on ScalePool, normalized to each respective baseline. ScalePool achieves an average speedup of 1.22$\times$ and up to 1.84$\times$ compared to the baseline. To analyze the contributing factors to this performance gain, we decompose execution times into three categories: communication time, computation time, and other time. Specifically, tensor parallelism communication within clusters occurs through NVLink, whereas pipeline and data parallelism communications across clusters utilize InfiniBand or CXL. Computation time represents the sum of GPU execution times for forward pass, backward pass, and optimizer steps. The other time category, relatively consistent across configurations, includes pipeline bubble and offloading overheads after excluding communication and computation times from the total execution time.

Breakdown analysis reveals that the performance gains predominantly result from reduced communication time. This improvement originates primarily from avoiding the scale-out domain's use of long-distance networks such as InfiniBand. InfiniBand-based RDMA communications inherently incur significant software overheads, including synchronization across communicators, which lead to longer latency. In contrast, CXL leverages hardware-based communication mechanisms, effectively eliminating software overheads and substantially reducing inter-cluster communication latency. Consequently, communication time across clusters is significantly decreased, achieving an average speedup of 3.79$\times$. In addition, the reduced pipeline parallelism communication time marginally decreases pipeline bubble durations, further contributing to the overall execution time speedup.

\begin{figure}
    \centering
    \includegraphics[width=\linewidth]{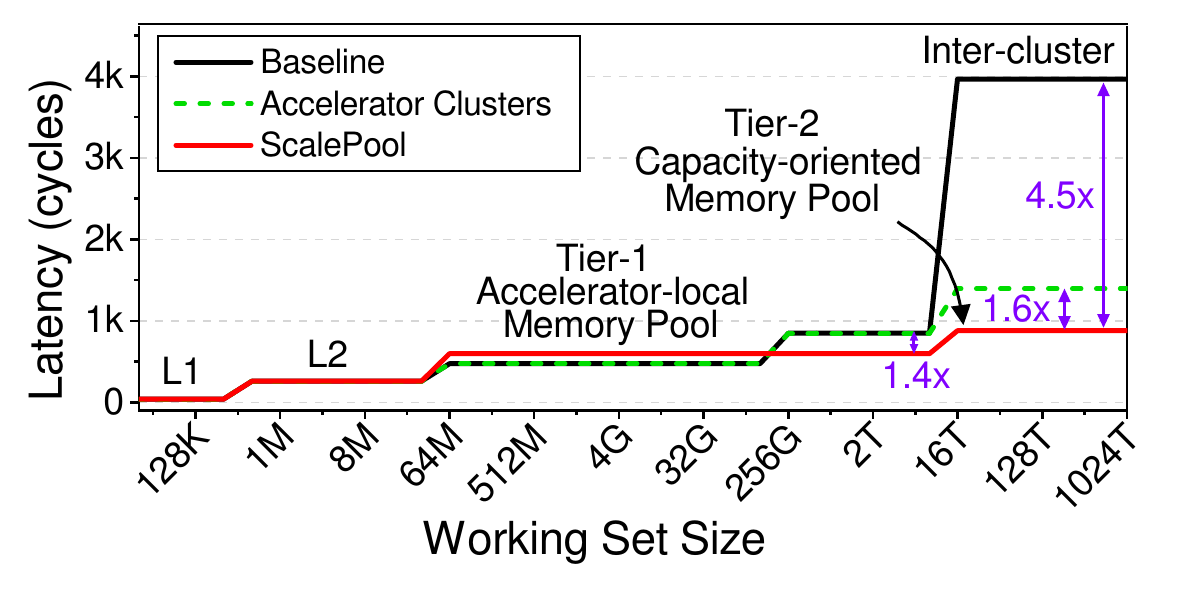}
    \vspace{-10pt}
    \caption{Performance results for tiered memory hierarchy.} \label{fig:eval_sensitivity}
\end{figure}

Figure~\ref{fig:eval_sensitivity} shows the latency impact of applying the tiered memory hierarchy within ScalePool across various working set sizes. Three configurations were evaluated sequentially: baseline, accelerator clusters, and tiered memory. The baseline configuration is identical to the architecture described in the previous evaluation. The accelerator clusters configuration refers to the accelerator-centric architecture, interconnected solely through CXL, while the tiered memory configuration further includes intra-cluster CXL connections to form a two-tier memory pool, comprising the ScalePool architecture.

When the working set exceeds the memory capacity of an individual accelerator, ScalePool achieves a 1.4$\times$ speedup compared to the baseline and accelerator clusters configurations. When the working set further exceeds the total memory capacity of an entire cluster, ScalePool attains a speedup of 4.5$\times$ and 1.6$\times$ over the baseline and accelerator clusters, respectively. These gains arise from resource disaggregation facilitated by the tiered memory hierarchy, enabling memory expansion through dedicated memory pools rather than relying exclusively on scaling cluster-level memory.

In cases where the memory requirement exceeds the accelerator's local memory capacity, the baseline and accelerator clusters configurations rely on non-coherent XLink-based access to other accelerators within the cluster. In contrast, ScalePool benefits from coherent, CXL-based tier-1 memory pools, resulting in improved performance. Furthermore, when memory requirements surpass the entire cluster's memory capacity, ScalePool leverages the tier-2 memory pool. This approach offers performance advantages compared to the RDMA-based baseline or the combined CXL and NVLink-based access mechanisms utilized in clusters.

\section{Acknowledgment}
The authors thank anonymous reviewers for their constructive feedback. This work is supported in part by Institute of Information \& communications Technology Planning \& Evaluation (IITP) grant funded by the Korea government (MSIT) (No.RS-2025-02214652, Development of SoC Technology for AI Semiconductor-Converged Pooled Storage/Memory, 20\%, No.RS-2025-02214654, Development of AI Semiconductor Converged Computational Memory/Storage SoC and Application Technology, 20\%, No.RS-2023-00221040, Enabling A Low-Power and High-Performance Computational SSD System that Supports the Execution of General-Purpose Applications), Ministry of Trade, Industry and Energy (MOTIE) and Korea Institute for Advancement of Technology (KIAT) through the International Cooperative R\&D program (No. P0028225), Korea Institute for Advancement of Technology (KIAT) grant funded by the Korea Government (MOTIE) (No. P0027923), and Technology development Program of MSS (RS-2023-00303967). This work is protected by one or more patents. Myoungsoo Jung is the corresponding author (mj@panmnesia.com).

\section{Conclusion}
We introduced a hybrid ScalePool architecture integrating accelerator-centric XLink and scalable CXL-based fabrics, accommodating diverse AI workloads. Our proposed tiered memory hierarchy and a customization of CXL balance high-performance local memory access with capacity-oriented memory pooling. Future work will evaluate this architecture in large-scale and explore additional optimizations.

\balance
\bibliographystyle{ACM-Reference-Format}
\bibliography{references}

\end{document}